\documentclass[lettersize,journal]{IEEEtran}
\usepackage{amsmath,amsfonts,amssymb, amsthm}
\usepackage{algorithmic}
\usepackage{algorithm}
\usepackage{array}
\usepackage[caption=false,font=normalsize,labelfont=sf,textfont=sf]{subfig}
\usepackage{textcomp}
\usepackage{stfloats}
\usepackage{url}
\usepackage{verbatim}
\usepackage{graphicx}
\usepackage{cite}
\usepackage{dsfont}

\usepackage{color}
\usepackage{cancel}
\usepackage{hyperref}

\newtheorem{example}{Example}

\newcommand{\be}{\begin{equation}}
\newcommand{\ee}{\end{equation}}

\newcommand{\al}{\alpha}
\newcommand{\tta}{\theta}
\newcommand{\E}{{\mathbb E}}
\newcommand{\cE}{{\cal E}}
\newcommand{\cH}{{\cal H}}
\newcommand{\N}{{\mathbb N}}
\newcommand{\si}{\sigma}
\newcommand{\nn}{\nonumber}
\newcommand{\q}{\quad}
\newcommand{\R}{{\mathbb R}}

\DeclareMathOperator*{\argmax}{argmax}

 \definecolor{darkspringgreen}{rgb}{0.09, 0.45, 0.27} 
 \definecolor{darkgray}{rgb}{0.66, 0.66, 0.66}

\begin{document}

\title{Node classification in networks via simplicial interactions}

\author{Eunho Koo and Tongseok Lim
\thanks{Eunho Koo (First Author): Chonnam National University, Gwangju 61186, Republic of Korea (kooeunho@jnu.ac.kr) 
\\
Tongseok Lim (Corresponding Author): Mitch Daniels School of Business, Purdue University, West Lafayette, Indiana 47907, USA (lim336@purdue.edu)}
}


\maketitle

\begin{abstract}
In the node classification task, it is natural to presume that densely connected nodes tend to exhibit similar attributes. Given this, it is crucial to first define what constitutes a dense connection and to develop a reliable mathematical tool for assessing node cohesiveness. In this paper, we propose a probability-based objective function for semi-supervised node classification that takes advantage of higher-order networks' capabilities. The proposed function reflects the philosophy aligned with the intuition behind classifying within higher order networks, as it is designed to reduce the likelihood of nodes interconnected through higher-order networks bearing different labels. Additionally, we propose the Stochastic Block Tensor Model (SBTM) as a graph generation model designed specifically to address a significant limitation of the traditional stochastic block model, which does not adequately represent the distribution of higher-order structures in real networks. We evaluate the objective function using networks generated by the SBTM, which include both balanced and imbalanced scenarios. Furthermore, we present an approach that integrates the objective function with graph neural network (GNN)-based semi-supervised node classification methodologies, aiming for additional performance gains. Our results demonstrate that in challenging classification scenarios—characterized by a low probability of homo-connections, a high probability of hetero-connections, and limited prior node information—models based on the higher-order network outperform pairwise interaction-based models. Furthermore, experimental results suggest that integrating our proposed objective function with existing GNN-based node classification approaches enhances classification performance by efficiently learning higher-order structures distributed in the network.
 
\end{abstract}


\section{Introduction}

Networks represented by graphs consist of nodes representing entities of the system, and edges depicting their interactions. Such graphical representations facilitate insights into the system’s modular structure or its inherent communities \cite{fortunato2016community, schaub2017many}. While traditional graph analysis methods only considered pairwise interaction between nodes, recent researches, including those in social sciences  \cite{easley2010networks} and biochemical systems \cite{klamt2009hypergraphs}, have experimentally demonstrated that networks in real systems often rely on interactions involving more than two nodes or agents. As a result, to analyze the attributes of a network, it is essential to illuminate the causal interactions of the network using higher-order networks (or hypergraphs) beyond pairwise relationships \cite{battiston2020networks}. There are various approaches to address this point of view, and recent studies are elucidating the relationships between cliques (a subset of nodes such that every two distinct nodes in the clique are adjacent) that form higher-order networks using probabilistic modeling based on the Stochastic Block Model (SBM) \cite{holland1983stochastic, ghoshdastidar2014consistency, kim2017community}. SBM is a generative model for random graphs that includes the following parameters: the number of nodes, the number of disjoint communities to which each node belongs, and the probability of edge connections between each community. The most common form of SBM assumes that the number of nodes in each community and the probability of edge connections within the same community are equal; nevertheless, several modified variants of SBM have also been studied \cite{karrer2011stochastic, airoldi2008mixed}. 

Studies related to community detection (or network clustering), on the other hand, have also been actively pursued \cite{blondel2008fast}. The goal of these studies is to divide the entire system’s nodes into several communities, with nodes in each community being densely connected internally \cite{newman2006modularity}. Research involving the higher-order network analysis has also advanced in this field, including the Bayesian framework \cite{vazquez2009finding}, $d$-wise hypergraph SBM on the probability of a hyperedge \cite{chien2018community}, the sum-of-squares method of SBM \cite{kim2017community}, and spectral analysis based on the Planted Partition Model (PPM), a variant of SBM \cite{ghoshdastidar2017consistency, ghoshdastidar2014consistency}. Notably, many studies solely evaluate the network's internal topology and do not account for prior information. However, in many real-world networks, even if it is a small part of the overall number of nodes, prior information is available; that is, we can use some known node labels as well as the total number of labels (communities). It has been reported that with only limited prior information, prediction accuracy and robustness in real noisy networks can be significantly improved \cite{eaton2012spin, ma2010semi}, and various methods have been suggested, including discrete potential theory method \cite{liu2014semi}, spin-glass model in statistical physics application \cite{eaton2012spin}, strategies integrating known cluster assignments for a fraction of nodes \cite{nowicki2001estimation}, and nonnegative matrix factorization model \cite{ma2010semi}. Recently, the integration of semi-supervised learning with Graph Neural Network (GNN) has significantly enhanced performance of node classification on large datasets. Graph Convolutional Network (GCN) \cite{kipf2017semisupervised} generates new node representations by aggregating features from nodes and their neighbors. Graph-SAGE \cite{hamilton2017inductive} introduces an inductive learning approach using node sampling and aggregation. Graph Attention Network (GAT) \cite{velikovi2018graph} apply attention mechanisms to neighboring nodes, and Jumping Knowledge Network (JKNet) \cite{xu2018representation} and Motif Graph Neural Network \cite{chen2023motif} collect information from an extended neighborhood range.

In this study, we propose a novel probability-based objective (loss) function for the semi-supervised node classification (community detection) task using higher-order networks. The loss function is motivated by the intuition that nodes densely interconnected with edges in a given network are likely to exhibit similar labels. It is intended to incentivize nodes in a hyperedge (a clique) to have the same label by imposing a natural penalty when nodes within the hyperedge have diverse labels. It is worth noting that the intuition is consistent with SBM’s general assumption that nodes with the same label are more likely to be connected in a network. In conjunction with the objective function, we use discrete potential theory to initialize the node probability distribution, specifically the solution to an appropriate Dirichlet boundary value problem on graphs, which can be effectively solved using the concept of equilibrium measures \cite{bendito2003solving}. 

We also propose a novel graph generation model, Stochastic Block Tensor Model (SBTM). In general, traditional SBM-generated networks differ significantly from many real-world networks. Specifically, when comparing networks of equivalent density (that is, networks with an identical count of nodes and edges), SBM-based models typically exhibit far fewer higher-order polyhedrons (simplices or cliques), than what is observed in real-world graphs. This limitation of SBM stems from its nature as an edge-generation model. For example, in social networks, while two people might form a friendship, it is also possible for three or more individuals to simultaneously establish a friendship. In light of this, we suggest that edge-generation models (SBM) have limits in producing network data that is similar to what is observed in the real world, and we offer a revised model capable of incorporating higher-order structures such as triangles and tetrahedrons into the network. 

Finally, we propose a method that achieves additional performance enhancements by integrating our proposed objective function with state-of-the-art (SOTA) GNN-based semi-supervised learning techniques. To the best of the authors' knowledge, many SOTA algorithms do not distinctively utilize all higher-order structures within graphs, but often focusing primarily on edges (pairwise interactions) instead. We explore how combining the strengths of both approaches can lead to further performance gains, and we validate this synergy on real citation network datasets (Cora, CiteSeer, and PubMed).
 
Our contributions are as follows:

$\bullet$ We propose a novel probability-based objective function for the semi-supervised node classification task, designed to leverage the full higher-order network structures.

$\bullet$ We extend SBM and introduce SBTM, a network generation model designed to simulate more realistic networks by incorporating higher-order structures.

$\bullet$ We explore and validate an approach for improving performance by combining the most recent graph neural network-based node classification methods with the objective function suggested in this paper.

This paper is structured as follows. Some preliminary information is provided in Section \ref{preliminaries}. The objective function is discussed in Section \ref{model}. The experimental setup is described in Section \ref{setup}. The result is evaluated in Section \ref{results}. Finally, in Section \ref{conclusion}, the conclusion and future work are presented.

\section{Preliminaries}\label{preliminaries}
In this section, we describe the preliminary graph node classification approach based on discrete potential theory after introducing some essential mathematical concepts.

\subsection{Higher-order networks}
In many real-world systems, network interactions are not only pairwise, but involve the joint non-linear couplings of more than two nodes \cite{bick2023higher}. Here, we fix some terminology on higher-order networks that will be used throughout the paper.

A (undirected) graph $G=(V,E)$ consists of a set $V=\{1,2,...,n\}$ of $n$ nodes, and a set $E \subset \{(i,j) \, | \, i,j \in V,\, i \ne j\}$ of edges. We have $(i,j) = (j,i) = \{i,j\}$ as $G$ is undirected. We assume that a graph $G$ is connected.\footnote{Because our proposed algorithm can be applied to each connected component of a graph, the assumption can be made without loss of generality.}

A hypergraph generalizes $E$ as $\cE \subset 2^V$ where $2^V$ denotes the power set of $V$, and we denote the hypergraph as $\cH=(V, \cE)$. In this paper, we focus on the case where $\cE$ consists of the {\em simplices} in $G$: For $k \in \N \cup \{0\}$ (the set of non-negative integers), a subset $\si =\{n_0,n_1,...,n_k\}$ of $V$ is called a $k$-simplex (which is  also called a $(k+1)$-clique) if the vertices $n_i \in V$ are distinct (i.e., $| \si | = k+1$) and for every $0 \le i < j \le k$, we have  $(n_i, n_j) \in E$. Let $K_k = E_{k-1}$ denote the set of all $k$-cliques, or $(k-1)$-simplices, in $G$. Note that $E_0 = V$, $E_1 = E$; a node is a 0-simplex, an edge a 1-simplex, a triangle a 2-simplex, a tetrahedron a 3-simplex, and so on. The set comprising all cliques of a graph $G$,
\be
K(G) := \bigcup_{k=1}^{\omega(G)}K_k(G),
\ee
is called the {\em clique complex} of the graph $G$. The clique number $\omega (G)$ is the number of vertices in a largest clique of $G$ \cite{lim2020hodge}. In this paper, we will consider $\cE$ a subset of $K(G)$ in terms of a hypergraph. 

\begin{example}
For a given $V=\{1,2,3,4\}$, consider $\cE = \big\{\{1\}, \{2\}, \{3\}, \{4\}, \{1, 2\}, \{1, 3\}, \{2, 3\}, \{3, 4\}, \{1, 2, 3\}\big\}$. 
\begin{figure}[H]
\centering
\includegraphics[width=2.0in]{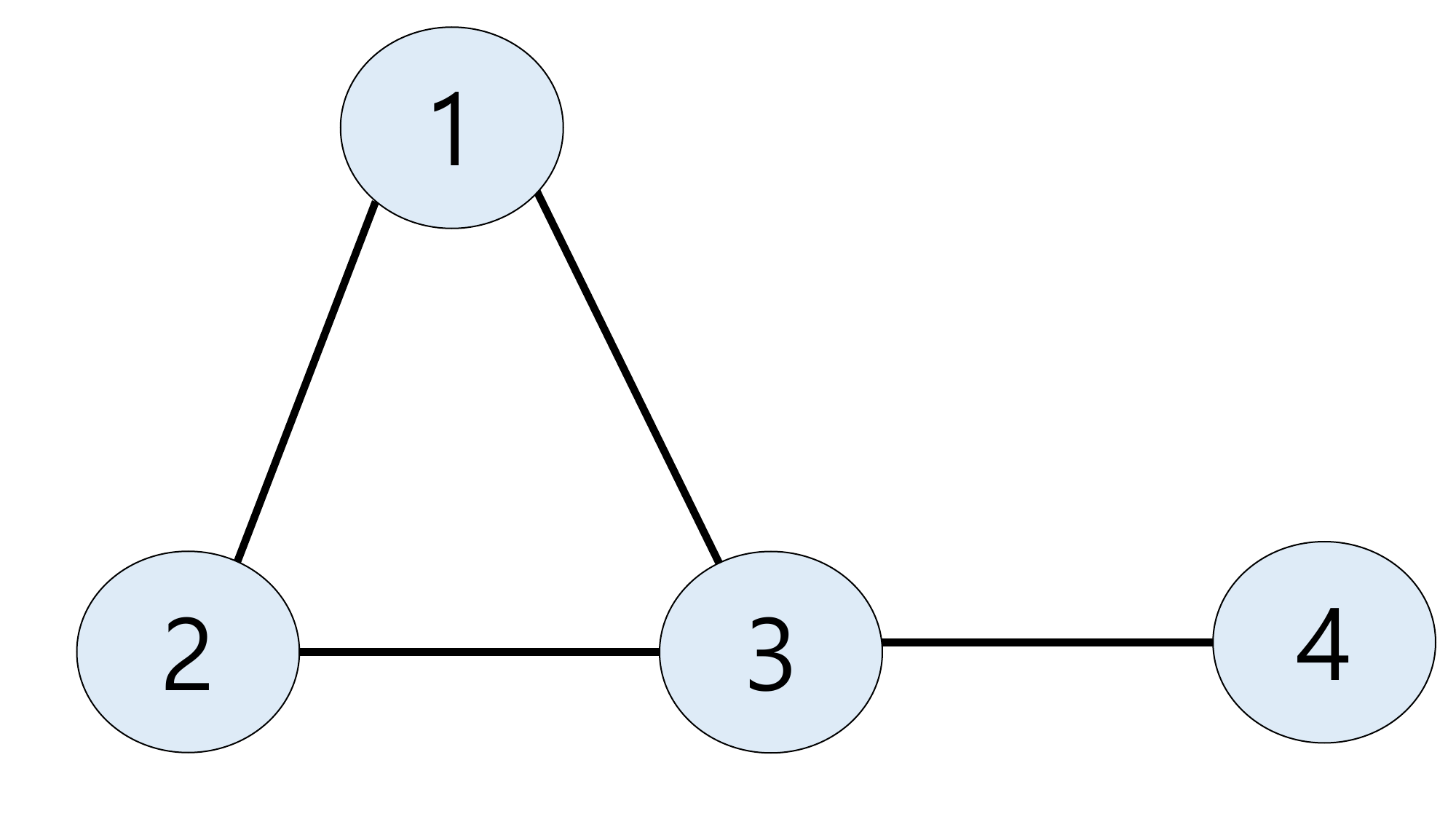}
\end{figure}
There are four $0$-simplices: $K_1 = \{\{1\}, \{2\}, \{3\}, \{4\}\}$; four $1$-simplices: $K_2 = \{\{1, 2\}, \{1, 3\}, \{2, 3\}, \{3, 4\}\}$; and one $2$-simplex: $K_3 = \{\{1,2,3\}\}$.
\end{example}

\subsection{Classification algorithm based on random walk on graphs
}\label{RW}

 In the semi-supervised node classification (partially labeled data classification) tasks, a classic and widely used algorithm based on a random walk on a graph  is the following. Given a graph $G=(V,E)$, a (unbiased) random walk moves from $i \in V$ to $j \in V$ with probability $1/k$ if $(i,j) \in E$ (i.e., $i$ and $j$ are adjacent) and the degree of $i$ (the number of nodes adjacent to $i$) is $k$. For a node set $V=\{1,2,\dots,n\}$ and a label-index set $I =\{1,\dots,l\}$, we assume that each node corresponds to one label in $I$ and we know the labels for only a small proportion of nodes relative to $|V| = n$. For $i \in I$ and $y \in V$, let $P_i (y)$ denote the probability that a random walk starting from an unlabeled node $y$ will reach an $i$-labeled node before arriving at any other labeled nodes. If $\argmax_{i\in I} P_i (y)=k$, the algorithm predicts that the label of the unlabeled node $y$ is $k$. If a node $y$ is already labeled as $k$, we have $P_i (y) = 1$ if $i = k$ and $P_i (y) = 0$ if $i \ne k$. In this study, we refer to this algorithm as \textsf{RW}, which stands for Random Walk.

 Now the question is how to obtain $P_i (y)$ for all $y \in V$ and $i \in I$. Potential theory shows $P_i (y)$ can be obtained from the solution $u$ of the following Dirichlet boundary value problem
\begin{align}\label{Dirichlet}
Lu(x) &= 0 \ \text{ if } \ x \in F = (E_i \cup H_i)^c, \\
u(x) &= 1 \ \text{ if } \ x \in E_i, \nn \\
u(x) &= 0 \ \text{ if } \ x \in H_i, \nn
\end{align}
where $L = D - A$ is the graph Laplacian matrix where $D,A$ are degree and adjacency matrix of a given graph $G$ (see \cite{lim2020hodge}), $E_i$ is the set of $i$-labeled nodes, $H_i$ is the set of labeled nodes excluding $i$-labeled nodes, and $u$ is a function on $V$, valued in $[0,1]$. Then it holds $P_i (y)=u(y)$ for all $y \in V$. 

Bendito, Carmona and Encinas \cite{bendito2003solving} proposed an elegant solution to the Dirichlet problem \eqref{Dirichlet} in terms of {\em equilibrium measures}. For any decomposition $V = F \cup F^c$ where $F$ and $F^c$ are both non-empty, they showed that there exists a unique measure (function) such that $Lv(x)= 1$ (and $v(x) > 0$) for all $x \in F$ and $Lv(x)=0$ (and $v(x)=0$) for all $x \in F^c$. The measure is called the equilibrium measure and denoted by $v^F$. Now for $V=F \cup F^c$ where $F,F^c$ are the set of unlabeled and labeled nodes, the solution $u$ of \eqref{Dirichlet} can be represented by 
\be\label{Dirichletsol}
u (x) = \sum_{z \in E_i} \frac{v^{ \{z\} \cup F}(x) - v^{F}(x)}{v^{ \{z\} \cup F}(z)}, \q x \in V.
\ee
Because $v^F$ can be obtained by solving a linear program, \eqref{Dirichletsol} provides an efficient way to solve the Dirichlet problem \eqref{Dirichlet}; see \cite{bendito2003solving} for more details.

We will use \textsf{RW} as a baseline algorithm for the semi-supervised node classification. Note that \textsf{RW} employs random walks and does not utilize  higher-order interactions (HOI). However, \textsf{RW} will be useful not only for comparing performance with our HOI-applied strategies, but also for providing a useful initialization method for training HOI algorithms. 

\section{Proposed model}\label{model}

\subsection{Simplicial objective function}

 \begin{figure*}[!t]
\centering
\includegraphics[width=6.0in]{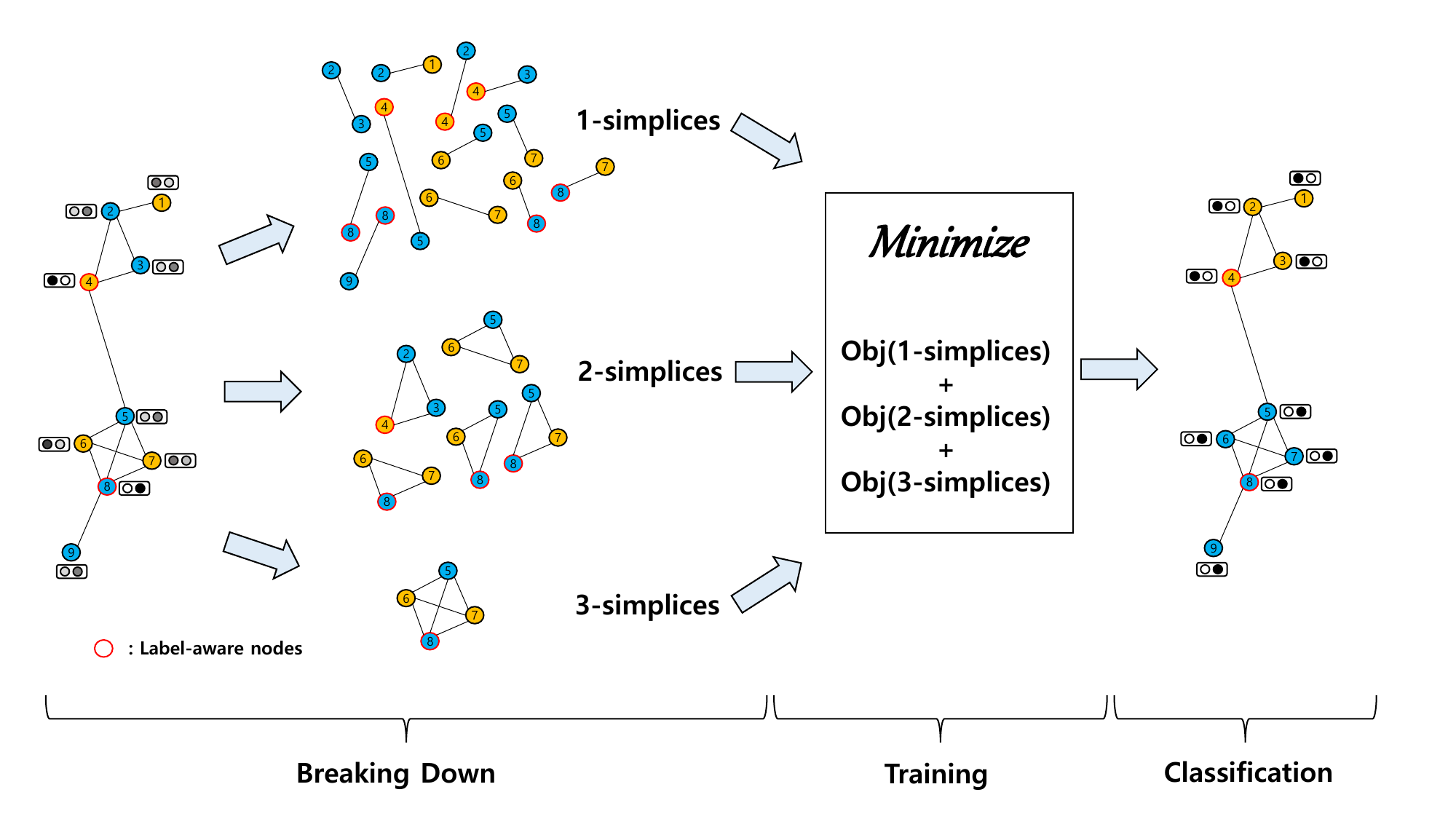}
\caption{Overall optimization process. First, we classify $k$-simplices (for $k=1,2,…$) in a given network based on their sizes (Breaking Down). Then, we multiply each size category by a distinct multinomial coefficient, sum these weighted values, and use this combined information to train a model (Training). Finally, we evaluate the model's classification performance (Classification).}
\label{process}
\end{figure*}

We propose a novel objective function for node classification which utilizes higher-order networks. Given a graph $G$, let $V=\{1,\dots,n\}$ denote the node set, $I=\{1,\dots,l \}$ denote the label index set, so the graph consists of $n$ nodes, and each node $j \in V$ has a label $i \in I$. Probability distribution over the labels for the node $j$ is denoted by $(p_1^j,p_2^j,...,p_l^j)$, where $p_i^j$ denotes the probability that node $j$ having label $i$, thus $\sum_{i=1}^lp_i^j=1$ for every $j \in V$. We define $K_k$ as the set of $(k-1)$-simplices in the graph, e.g., $K_1, K_2, K_3$ correspond to the set of nodes, edges, triangles, respectively. Let $M = \omega(G)$ be the maximum possible value of $k$, that is, the simplex composed of the most nodes in the graph is a $(M-1)$-simplex with $M$ nodes. Now we define an objective function  for node classification task by
\begin{align}\label{objective}
J = \sum_{k=2}^M w_k \sum_{(j_1,...,j_k) \in K_k} \sum_{(i_1,...,i_k)=\tta \in  I^k} C_\tta p_{i_1}^{j_1}  p_{i_2}^{j_2}\dots  p_{i_k}^{j_k}
\end{align}
where $I^k = I \times \dots \times I$ ($k$ times), $w_k \ge 0$ are weight hyperparameters, and $C_\tta := {k \choose e_1,e_2,...,e_l} = \frac{k!}{e_1!e_2!\dots e_l!}$ where $e_i$ is the number of occurrences of the label $i$ in $(i_1,i_2,…,i_k )=\tta$ with $\sum_{i=1}^l e_i=k$. Hence $J$ is determined by the underlying graph structure, and is a function of probabilities $\{p_i^j\}_{i \in I, j \in V}$. 

We then proceed to solve the minimization problem:
\begin{align}\label{problem}
\text{Minimize } \, J \, \text{ over }  \, \Delta^n = \Delta \times \Delta \times \dots \times \Delta
\end{align}
where $\Delta$ is the probability simplex in $\R^l$, such that $p^j := (p_1^j, p_2^j,...,p_l^j) \in \Delta$. 

The idea is that a higher penalty is assigned via a multinomial coefficient $C_\tta$ to the simplex with a greater diversity of labels and vice versa, and we sum the penalties over all $(k-1)$-simplices, finally taking a weighted sum over $K_k$. This leads us to expect that a probability distribution $\{p_i^j\}_{i \in I, j \in V}$ minimizing $J$ over $\Delta^n$ will find a node labeling that encourages the least diversity of labels within each simplex on average. This is consistent with our model assumption that the connection probability within the same label is higher than between different labels in forming the network. Finally, from a computational standpoint, solving the problem \eqref{problem} may necessitate a suitable initialization of the value $\{p_i^j\} \in \Delta^n$. We employ \textsf{RW} and use its solution as our initial value, which is simple to compute through linear programs. This is the main idea of this paper and is illustrated in Figure \ref{process}.

In this paper, we use an exponential base weight $w_k=\al^{k-1}$ in \eqref{objective} with various base values $\al >0$, and the default value of $\al$ is set to $1$ leading to the constant weight. It is worth noting that the expected number of $k$-cliques can be shown to be bounded by $ \E [| K_k |] \le 2^{k-1} |V|^k p^{\frac{k(k-1)}{2}}$, if the underlying network is generated by a stochastic block model with an edge-probability matrix with probability $p$ on the diagonal and $q$ off-diagonal, where $p>q$. This explains why, contrast to the many real-world network datasets, higher-order simplicial structures become increasingly difficult to observe in SBM as $k$ increases, due to the above bounding estimate with small $p$.

\begin{example}\label{ex2}
 Let  $I=\{1,2\}$ and let $K_3$ be the set of all $2$-simplices in a given graph. Fix a member $(j_1,j_2,j_3 ) \in K_3$. Then possible combinations for the binary labeling leads to $2^3=8$ terms of the form $p_{i_1}^{j_1} p_{i_2}^{j_2} p_{i_3}^{j_3}$ where $i_1, i_2, i_3 \in I$, which represents the probability that the nodes $j_1,j_2,j_3$ have labels $i_1,i_2,i_3$, respectively. We impose a penalty of multinomial order according to the distinct node labels within a given simplex. For a $2$-simplex, the objective with respect to two labels classification, which is the third summation term in \eqref{objective}, consists of the following eight terms: 
\begin{align*}
&{3 \choose 3,0} p_1^{j_1}p_1^{j_2}p_1^{j_3} + {3 \choose 2,1} p_1^{j_1}p_1^{j_2}p_2^{j_3} + {3 \choose 2,1} p_1^{j_1}p_2^{j_2}p_1^{j_3}\\
&+  {3 \choose 2,1} p_2^{j_1}p_1^{j_2}p_1^{j_3} 
+ {3 \choose 1,2} p_1^{j_1}p_2^{j_2}p_2^{j_3} + {3 \choose 1,2} p_2^{j_1}p_1^{j_2}p_2^{j_3}\\
& + {3 \choose 1,2} p_2^{j_1}p_2^{j_2}p_1^{j_3} + {3 \choose 0,3} p_2^{j_1}p_2^{j_2}p_2^{j_3}, 
\end{align*}
where ${3 \choose 3,0} = {3 \choose 0,3} = 1$, ${3 \choose 2,1} = {3 \choose 1,2} = 3$, and $p_1^j + p_2^j = 1$.
\end{example}
As can be inferred from Example \ref{ex2}, in the computation related to the edge set $K_2$ in \eqref{objective}, the computational complexity is $O(|K_2 | 2^l)$ where $|K_2 |$ is the number of edges in the network, and $l$ is the number of labels. In general, for each $k$-clique set $K_k$, the computational complexity can be derived as $O(|K_k |k^l)$, and therefore, the total complexity for all simplices up to size $M=\omega(G)$ would be $O(\sum_{k=2}^M |K_k| k^l)$. Since the complexity depends on the number of cliques in the network and the number of labels, the complexity of the proposed objective function increases with the inclusion of more higher-order simplices and a greater number of distinct labels.

\subsection{SBTM: a new graph generation approach }

Initially conceptualized in social networks and bioinformatics, Stochastic Block Model (SBM) \cite{holland1983stochastic} harnesses a probabilistic approach, giving a simple generative model for random graphs. For a node set and a label set $V$ and $I$ as above, we consider $l = |I|$ distinct labels or communities, each being a non-empty disjoint subsets of $V$ for each $i \in I$. The connection probabilities between nodes in $V$ by edges is identified by a $l \times l$ edge-probability matrix $B$, where $B_{ij}$ indicates the probability that a node belonging to the $i$th label connects with a node in the $j$th label by an edge. In many cases, the diagonal elements of $B$ are greater than off-diagonal elements, implying that the connection probability within the same label is higher than between different labels.

However, SBM has notable limitations: it lacks strong local clustering, failing to capture dense substructures like triangles that are essential for representing complex network dynamics. Additionally, SBMs assume edge independence, which is unrealistic since real-world networks often exhibit interdependencies among edges along with a combination of hierarchical community structures \cite{paul2023higher}. When comparing networks of equal density—networks with an identical number of nodes and edges—SBM-based models generally yield far fewer high-dimensional polyhedrons (simplifies or cliques) than what is commonly observed in real-world networks. This shortcoming arises from SBM’s foundation as an edge-generation model, which limits its ability to form interconnected groups beyond pairs. For example, in collaboration networks where multiple individuals contribute to the same project or publication, groups of three or more people often establish simultaneous collaborative connections. Such higher-order connections are challenging for traditional SBMs to capture accurately. Consequently, edge-generation models like SBM struggle to recreate the complex, interdependent structures present in real-world networks. To address this, in this paper we present the Stochastic Block Tensor Model (SBTM), a natural extension of SBM by assigning probability tensors (rather than a single probability matrix) to generate $k$-cliques for each $k \ge 2$, facilitating the development of higher-order simplices in the network. This leads to the emergence of higher-order simplices in the synthetic network, resulting in networks that closely resemble real-world networks.

To explain the concept of SBTM, we recall that a SBM is described by its probability matrix $B \in \R^{l^2}$, whose $(i,j)$ entry $B_{ij}$ reflects the probability that nodes $i$ and $j$ are connected by an edge in a random graph $G$. Each edge between $i, j$ is generated independently in $G$ through $B_{ij}$. Motivated by this, for each $k \ge 2$, let $T_k \in \R^{l^k}$ be a probability tensor whose $(i_1,\dots, i_k)$ entry $(T_k)_{i_1,...,i_k}$ indicates the probability that the nodes $i_1,\dots, i_k$ are interconnected as a $k$-clique in $G$. Given probability tensors $T_2, T_3,...$, we assume that an SBTM generates a random graph $G$ as follows: for each $k \ge 2$, we generate $k$-cliques independently for every size-$k$ subset of $V$ using $T_k$. Let $G_k$ denote the graph that represents the (unweighted) overlap of all  randomly created $k$-cliques. The final graph $G$ is defined as the full overlap of all $G_k$. Note that an edge between two nodes $i,j$ can be generated more than once in this procedure. In this paper, we assume that $G$ is unweighted, meaning that an edge between $i,j$ exists in $G$ if and only if it is produced in at least one of $G_k$. However, one can also define a weighted graph by counting the number of edge generation between each pair of edges. As a result, SBM is a specific instance of SBTM, with $T_k \equiv 0$ for all $k \ge 3$.

Thus SBTM is given by a set of probability tensors $(T_k)_{k \ge 2}$. In this study, we will define these tensors as follows. Let $N=[N_1,\dots,N_l]$ denote the number of nodes for each label. For each $k \ge 2$, we choose a probability matrix $B_k \in \R^{l^2}$. $T_k$ will be defined using $B_k$. Given a sequence of labels $(i_1,...,i_k)=\tta \in I^k$, we define the $\tta$-entry of $T_k$, denoted by $(T_k)_\tta$, as 
\be\label{probytensor}
(T_k)_\tta := \prod_{i=1}^l { N_i \choose e_i} \prod_{1 \le a < b \le k} (B_k)_{i_a i_b},
\ee
where $e_i$ is the number of occurrences of label $i$ in $ (i_1,...,i_k)$. For example, in 4-clique generation, if $N=[N_1,N_2,N_3]$ and $\tta=(1,1,3,1)$, then $e_1=3, e_2=0, e_3=1$, and the entry at $(1,1,3,1)$ in the tensor $T_4$ is given by \eqref{probytensor} with $l=3$, $k=4$. 

 The formula \eqref{probytensor} is motivated from calculating the expected number of $k$-cliques in the traditional SBM: The first term $\prod_{i=1}^l { N_i \choose e_i}$ calculates the number of ways to select $e_i$ nodes from each group of $N_i$ nodes, and the second term $\prod_{1 \le a < b \le k} (B_k)_{i_a i_b}$ is the product of the connection probabilities provided by $B_k$ between nodes in a $k$-clique. The second term will be small if the clique has diversified labels because, intuitively, the diagonal entries of $B_k$ are larger than the off-diagonal entries, reflecting a higher probability of connections within the same cluster than between different clusters. By assigning appropriate probabilities to each element of $B_k$, we intend to systematically determine a probability value to each entry of $T_k$, where it takes lower values for diversified labels and vice versa for uniform labels.

\subsection{Related works on graph semi-supervised learning}
Graph semi-supervised learning (GSSL) is designed to utilize inexpensive unlabeled data to improve the model's effectiveness using only a limited number of expensive labeled data. Hence, the semi-supervised learning framework aligns well with numerous practical scenarios where acquiring labels is challenging. According to a taxonomy framework \cite{song2022graph}, GSSL can be categorized into graph regularization methods, matrix factorization-based methods, random walk-based methods, autoencoder-based methods, and GNN-based methods. 

The graph regularization method in GSSL involves adding a smoothness constraint to the prediction function, ensuring that nodes with similar features have similar labels, often using the graph Laplacian to enforce this constraint. Examples of graph regularization methods include Gaussian random field \cite{zhu2003semi}, local and global consistency \cite{zhou2003learning}, directed regularization \cite{zhou2005learning}, linear neighborhood propagation \cite{wang2006label}, and manifold regularization with cosine distance \cite{belkin2006manifold}. The matrix factorization-based method in GSSL involves factorizing a matrix that represents relationships between node pairs to obtain node embeddings, utilizing matrices like the adjacency matrix or the normalized Laplacian matrix to capture underlying graph structures. Notable works in this category include locally linear embedding \cite{roweis2000nonlinear}, Laplacian eigenmaps \cite{belkin2001laplacian}, GraRep \cite{cao2015grarep}, and HOPE \cite{ou2016asymmetric}. The random walk-based method in GSSL uses random walks to capture graph properties like node centrality and similarity. Key works in this category include DeepWalk \cite{perozzi2014deepwalk}, LINE \cite{tang2015line}, Planetoid \cite{yang2016revisiting}, node2vec \cite{grover2016node2vec}, and VG-GCN \cite{hong2021variational}. The autoencoder-based method in GSSL uses autoencoders to encode each node into a low-dimensional embedding while preserving graph structure. Significant autoencoder-based methods include structural deep network integration \cite{wang2016structural}, deep neural networks for learning graph representations \cite{cao2016deep}, graph autoencoder \cite{kipf2016variational}, and mGNN \cite{wang2022minority}. The GNN-based method in GSSL utilizes Graph Neural Networks to generate node embeddings by aggregating information from a node's neighbors. Prominent GNN-based methods include the graph neural network model \cite{scarselli2008graph}, graph convolutional network \cite{kipf2017semisupervised}, deep sets \cite{zaheer2017deep}, graph-SAGE \cite{hamilton2017inductive}, graph attention network \cite{velikovi2018graph}, jumping knowledge network \cite{xu2018representation}, simplifying graph convolutional networks \cite{pmlr-v97-wu19e}, Motif graph neural network (MGNN) \cite{chen2023motif}, GDPNet \cite{WangYWC0ZHC19}, Infograph \cite{SunHV020}, NeuralSparse \cite{pmlr-v119-zheng20d}, PTDNet \cite{10.1145/3437963.3441734}, and SPC-GNN \cite{gong2022self}.

\section{Experimental setup}\label{setup}

  \begin{table*}[!t]
\centering
\includegraphics[width=6.72in]{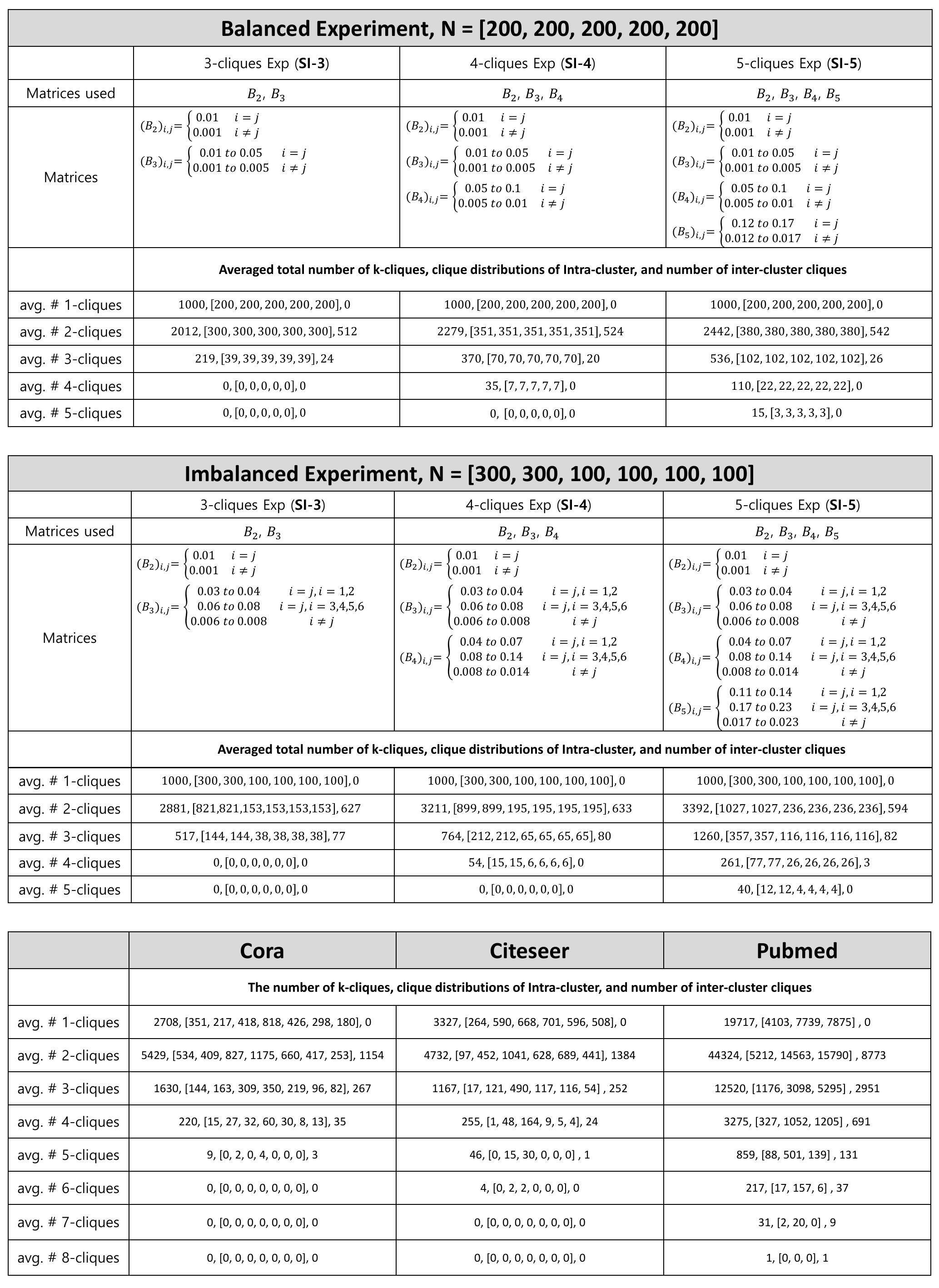}
\caption{Balanced and Imbalanced \textsf{SI-3}, \textsf{SI-4}, \textsf{SI-5} experiment settings, benchmark data, and clique distributions in each experiment.}
\label{cliquetable}
\end{table*}

In this study, we utilize SBTM to generate balanced and imbalanced graphs to evaluate the proposed objective function. The maximum clique size $M$ is set to 5 in each experiment. The balanced graph is set with $N=[200,200,200,200,200]$, meaning $|I| = 5$, $|V|=1000$, and the number of nodes corresponding to each label is 200. The imbalanced graph is set with $N=[300,300,100,100,100,100]$, thus $|I| = 6$ and $|V|=1000$. Additionally, the $k$-clique probability matrix $B_k \in \R^{l^2}$ for $k=2,3,4,5$ with different probability entries are chosen for both balanced and imbalanced networks. Detailed information is presented in Table \ref{cliquetable}.

The diagonal and off-diagonal probability settings in $(B_k)_k$ in both balanced and imbalanced graphs aim to investigate the relationship between performance improvement and variations in the number and distribution of higher-order simplices. The experimental setting ranges $k$ from 2 to 5, allowing the performance of the proposed objective function to be tested across a wide range of higher-order clique distributions.

 \begin{figure*}[!t]
\centering
\includegraphics[width=6.6in]{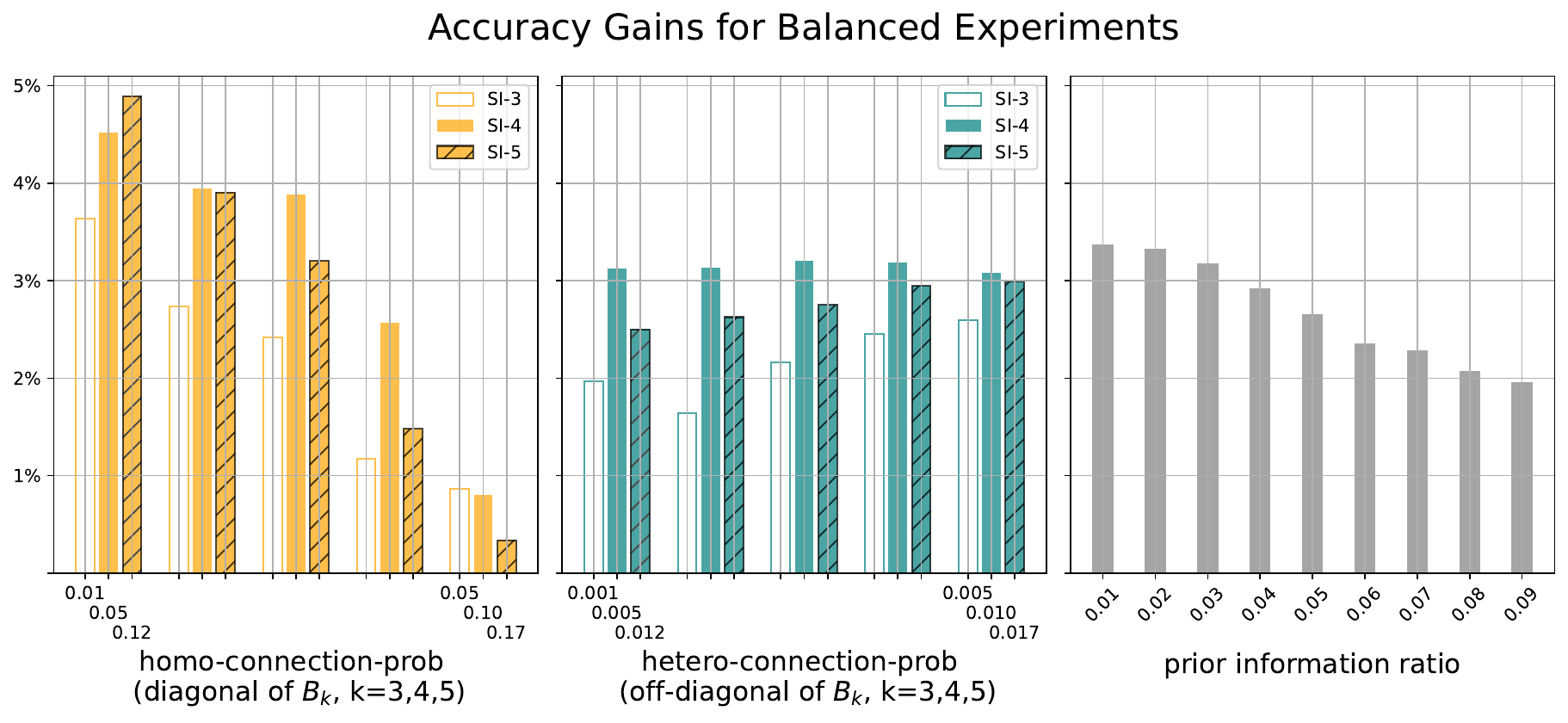}
\caption{Accuracy gains for balanced \textsf{SI-3}, \textsf{SI-4} and \textsf{SI-5} experiments with respect to variations in homo-connection probability (left), hetero-connection probability (middle), and prior information ratio in $N$=[200,200,200,200,200] setting. In left, diagonal of $B_3$, $B_4$, and $B_5$ varies from 0.01 to 0.05, from 0.05 to 0.10, and from 0.12 to 0.17, respectively. In middle, off-diagonal of $B_3$, $B_4$, and $B_5$ varies from 0.001 to 0.005, from 0.005 to 0.010, and from 0.012 to 0.017, respectively. In right, the performances are tested ranging from 10 known nodes (= prior information ratio 0.01) to 90 known nodes (PIR 0.09).}
\label{balancedfigure}
\end{figure*}

We define \textsf{SI-m} (Simplicial Interactions up to order $m$) as the experiment on the objective function utilizing up to $K_m$ (the set of $(m-1)$-simplices in the network) for $m=3,4,5$. More specifically, we define on the domain $\Delta^n$ the objective
\begin{align}
J_m = \sum_{k=2}^m w_k \sum_{(j_1,...,j_k) \in K_k} \sum_{(i_1,...,i_k)=\tta \in I^k} C_\tta p_{i_1}^{j_1}  p_{i_2}^{j_2}\dots  p_{i_k}^{j_k},\nn
\end{align}
noting that the sum is over $k=2$ to $m$, thereby employing up to $(m-1)$-simplices in the graph. In particular, \textsf{SI-2} utilizes only pairwise interactions through edges in the network, and the corresponding objective function becomes
\begin{align}
J_2 = \sum_{(j_1,j_2) \in K_2} \sum_{(i_1,i_2)=\tta \in I^2} C_\tta p_{i_1}^{j_1}  p_{i_2}^{j_2}. \nn
\end{align}
The experimental results show that the performances of \textsf{RW} and \textsf{SI-2} are comparable. Because of this, the average performance of these two methods in experiments that do not utilize higher-order networks will be presented as \textsf{PI} (Pairwise Interactions).  In this semi-supervised learning research, the prior information ratio (the proportion of nodes whose labels are revealed) is varied from 1\% to 9\%, and nodes corresponding to the given prior information ratio are randomly sampled for each cluster and used as label-aware nodes. The results are evaluated for the performance gains achieved by \textsf{SI-M} (where $M= \omega(G)$) over the \textsf{PI} for each prior information ratio. In this study, the accuracy gain is defined as the relative performance improvement by $y/x-1$ where $x$ and $y$ represents the accuracies obtained from \textsf{PI} and \textsf{SI-M}, respectively.

We adopt a transductive learning framework using three well-known citation network datasets: Cora, Citeseer, and Pubmed, as outlined by \cite{sen2008collective} and following the methodology of \cite{yang2016revisiting} and \cite{velikovi2018graph}. These datasets include the Cora dataset with 2708 nodes, 5429 undirected citation edges, and 7 classes; the Citeseer dataset comprises 3327 nodes, 4732 edges, and 6 classes; and the Pubmed dataset with 19717 nodes, 44338 edges, and 3 classes. In our experimental setup, documents are represented as nodes, and citations are represented as edges. The distribution of simplices in the benchmark dataset is illustrated in Table \ref{cliquetable}.

In the training process, Glorot initialization \cite{glorot2010understanding} is utilized to initialize the parameters, and the Adam SGD optimizer \cite{KingmaB14} is employed for optimization. For all experiments, the learning rate is set to 0.4 and the number of epochs to 10. The proposed objective function aims to learn the probability distribution assigned to each node in the network. We apply the softmax function to describe node probability distributions. At the end of the training process, the argmax function is used to obtain the final classification results. The performance is evaluated using the accuracy error metric.

 In this study, we introduce a simple and concise learning process that assigns an $l$-dimensional probability distribution to each node in a given network, categorizes the network's simplices by their size, and integrates them into objective function \eqref{objective} for training (Figure \ref{process}). Specifically, for a fixed weight hyperparameter $w_k$ in \eqref{objective}, this research utilizes only the learning rate and epochs as hyperparameters. As a result, the proposed algorithm (which does not rely on node embedding dimensions, number of layers in the neural network architecture, number of units per layer, dropout rates, or activation functions as hyperparameters) offers enhanced implementation efficiency compared to the existing GNN architectures that demand extensive hyperparameter tuning. Leveraging this efficiency, we evaluate the proposed objective function exclusively with training data (label-aware nodes) and test data (label-unaware nodes) in our experiments involving both synthetic graphs and real citation networks (Cora, Citeseer, Pubmed), without the use of validation nodes.

\section{Results}\label{results}

In this section, we evaluate the node classification performance of the proposed higher-order networks based objective function on balanced data (Section \ref{balanced}), and imbalanced data (Section \ref{imbalanced}). In both sections, we primarily evaluate the accuracy gain achieved by using the proposed objective function with higher-order structures, compared to the performance of \textsf{PI} (mean performance of \textsf{RW} and \textsf{SI-2}). Furthermore, we discuss how the objective function can be combined with various semi-supervised node classification methodologies to achieve additional performance gains (Section \ref{GNN}). Result summary and discussion is presented in Section \ref{summary}.  The implementation algorithm can be found at \url{https://github.com/kooeunho/HOI_objective}.

\subsection{Balanced experiment}\label{balanced}

 \begin{figure*}[!t]
\centering
\includegraphics[width=6.60in]{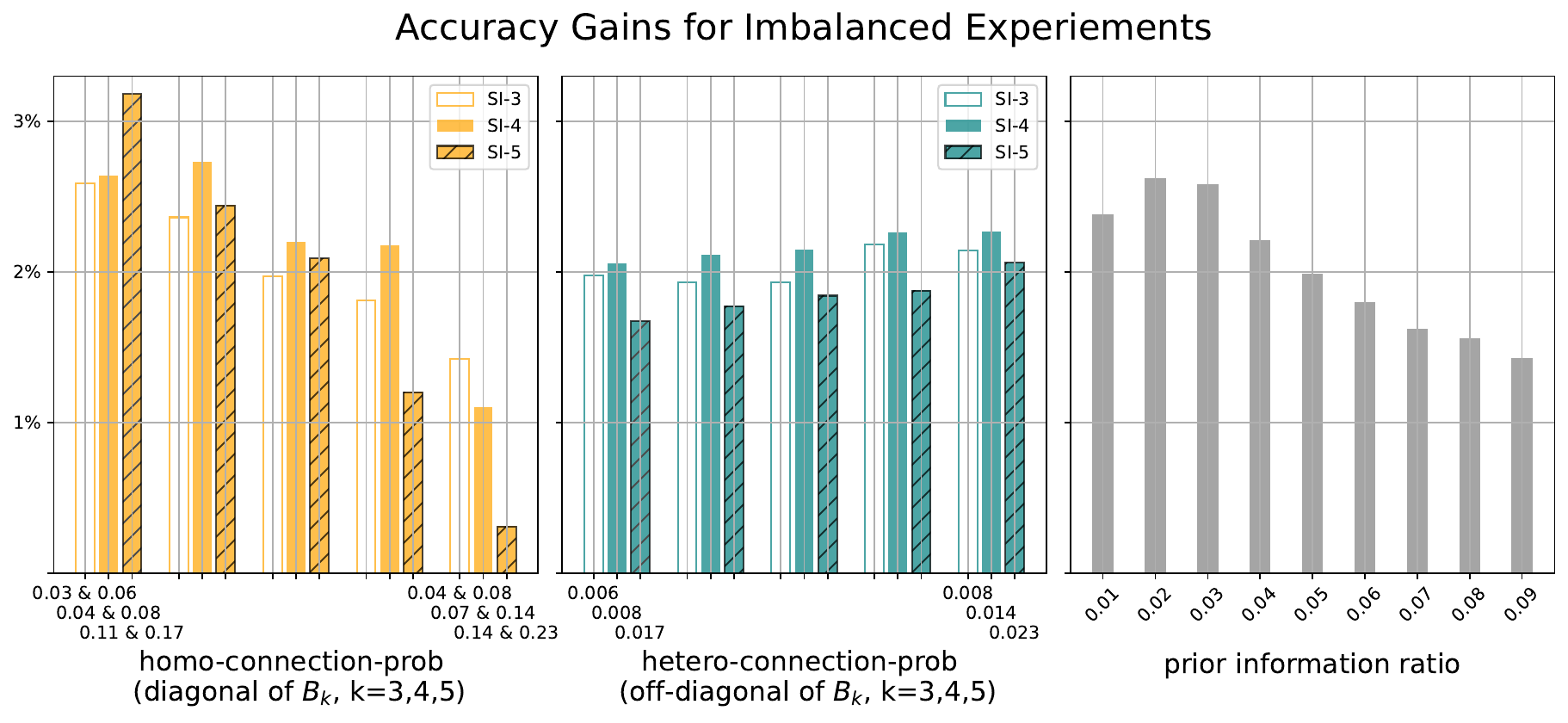}
\caption{Accuracy gains for imbalanced \textsf{SI-3}, \textsf{SI-4} and \textsf{SI-5} experiments with respect to variations in homo-connection probability (left), hetero-connection probability (middle), and prior information ratio in $N$=[300,300,100,100,100,100] setting. In left, diagonal of $B_3$, $B_4$ and $B_5$ varies from 0.03 \& 0.06 to 0.04 \& 0.08 (that is, the diagonal entries of $B_3$ varies from [0.03, 0.03, 0.06, 0.06, 0.06, 0.06] to [0.04, 0.04, 0.08, 0.08, 0.08, 0,08]), from 0.04 \& 0.08 to 0.07 \& 0.14, and from 0.11 \& 0.17 to 0.14 \& 0.23, respectively. In middle, off-diagonal of $B_3$, $B_4$ and $B_5$ varies from 0.006 to 0.008, from 0.008 to 0.014, and from 0.017 to 0.023. In right, the performances are tested ranging from 10 known nodes (= prior information ratio 0.01) to 90 known nodes (PIR 0.09).}
\label{imbalancedfigure}
\end{figure*}

We present the experimental results based on the balanced experiment settings depicted in Table \ref{cliquetable}. Three experiment settings (\textsf{SI-3}, \textsf{SI-4} and \textsf{SI-5}) are designed to generate higher-order simplices encompassing 3-cliques, 4-cliques, and 5-cliques, respectively. We compare the accuracy gains of our proposed objective function, which utilizes the higher-order simplices distributed across the network, with \textsf{PI}. Figure \ref{balancedfigure} illustrates the accuracy gains of our strategy relative to the \textsf{PI} with respect to the homo-connection probability (diagonals of $B_k$ for each $k=3,4,5$), the hetero-connection probability (off-diagonals of $B_k$ for each $k=3,4,5$), and the prior information ratio. The weight hyperparameter $w_k$ is set to $1$ in balanced experiments. The experimental results show that the accuracy gains for the \textsf{SI-3}, \textsf{SI-4} and \textsf{SI-5} experiments are 2.17\%, 3.15\%, and 2.76\%, respectively. This shows that the proposed objective function enhances performance by utilizing higher-dimensional simplices such as triangles, tetrahedrons and pentachorons. Furthermore, according to Figure \ref{balancedfigure}, greater accuracy increases occur with lower homo-connection probabilities, higher hetero-connection probabilities, and lower prior information ratios.  This implies that the objective function can achieve significant additional performance gains by exploiting higher-order structures in challenging scenarios where the network structure is ambiguous and prior information is limited.

\subsection{Imbalanced experiment}\label{imbalanced}

 \begin{figure*}[!t]
\centering
\includegraphics[width=6.60in]{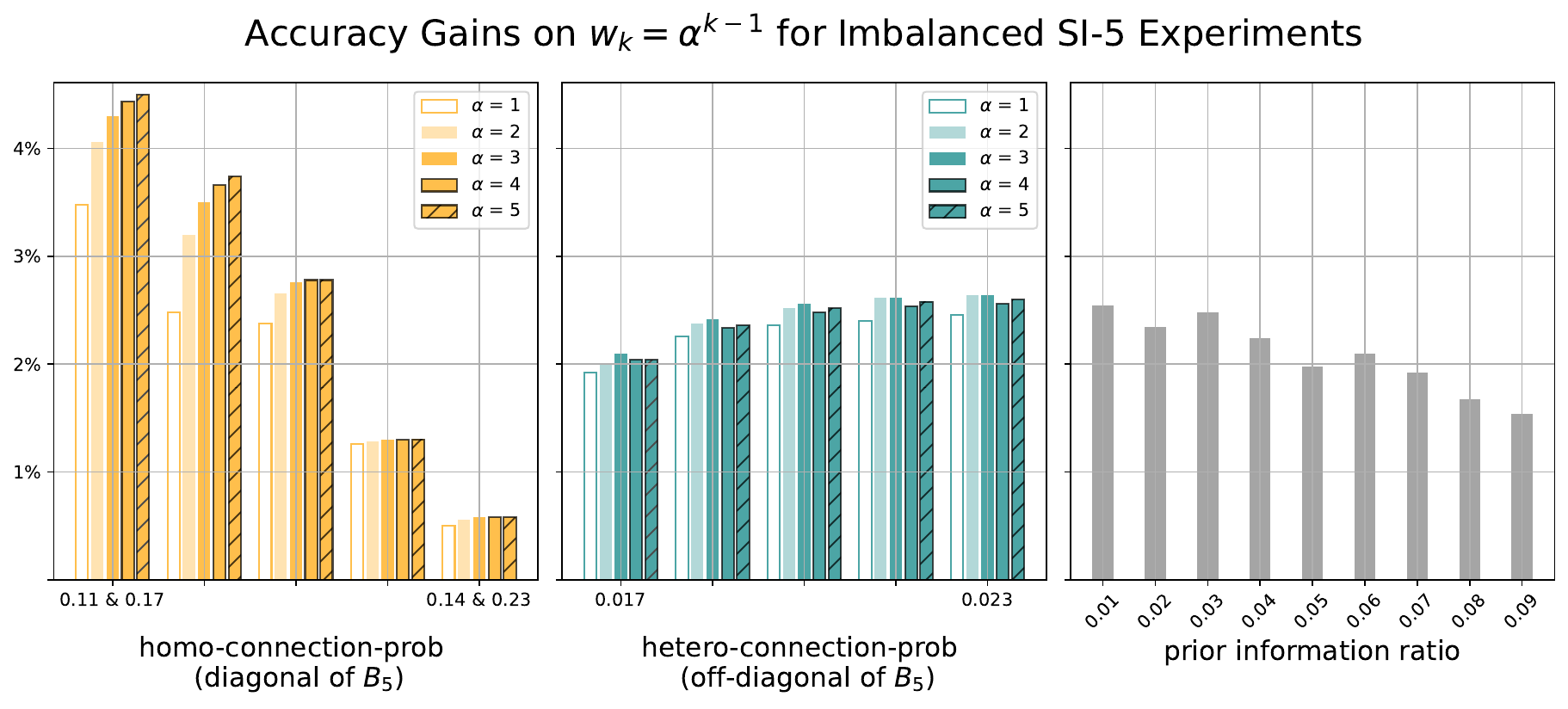}
\caption{Accuracy gains on $w_k=\alpha^{k-1}$ in the objective \eqref{objective} for imbalanced \textsf{SI-5} experiments with respect to variations in homo-connection probability (left), hetero-connection probability (middle), and prior information ratio in $N$=[300,300,100,100,100,100] setting. In left, diagonal of $B_5$ varies from [0.11, 0.11, 0.17, 0.17, 0.17, 0.17] to [0.14, 0.14, 0.23, 0.23, 0.23, 0,23]. In middle, off-diagonal of $B_5$ varies from 0.017 to 0.023. In right, the performances are tested ranging from 10 known nodes (= prior information ratio 0.01) to 90 known nodes (PIR 0.09).}
\label{weightfigure}
\end{figure*}

In the imbalanced experiments, the parameters of the SBTM (that is, $B_k$ for $k=3,4,5$) differ from those in the balanced experiments, as shown in Table \ref{cliquetable}. We adjust these parameters because our objective function aims to enhance node classification performance by leveraging the higher-order structures in the network, and a particular cluster containing too few of these structures would not allow for an accurate evaluation of the proposed function. The resulting distribution of $k$-cliques in each cluster corresponding to the set parameters is presented in Table \ref{cliquetable}. The experimental results show that the accuracy gains for the \textsf{SI-3}, \textsf{SI-4} and \textsf{SI-5} experiments against \textsf{PI} are 2.03\%, 2.17\%, and 1.84\%, respectively. This result implies that higher-order structures have a significant impact on node classification performance as in the balanced experiments. Additionally, the characteristic of the proposed objective function to achieve more performance gains in challenging scenarios is also evident in the imbalanced settings, as seen in Figure \ref{imbalancedfigure}. 
 
 The weight hyperparameter $w_k$ on $k$-cliques in the objective function \eqref{objective} is designed to emphasize higher-order simplices over lower-order ones. In the imbalanced experiments, additional tests employing 
$w_k=\alpha^{k-1}$ (with $\alpha=1,2,3,4,5$) on \textsf{SI-5}  experiments (Figure \ref{weightfigure}) show that mean accuracy gains for \textsf{SI-5}  against \textsf{PI} are 2.02\%, 2.35\%, 2.49\%, 2.55\%, and 2.58\%, for $\alpha=1,2,3,4$ and $5$, respectively. Greater weights on higher-order simplices lead to enhanced performance; however, incremental benefits diminish as $\alpha$ increases. Employing the weighting strategy that favors higher-order cliques is suggested to potentially improve node classification in imbalanced networks, including those with long-tail distributions.
 
\subsection{Integration of the objective function with GNN}\label{GNN}
 
 In this subsection, we propose a method to enhance node classification performance by integrating the objective function proposed in this study with existing GNN-based node classification methods, which do not directly utilize higher-order structures in training, including the latest semi-supervised and unsupervised learning techniques. Existing methodologies employ various strategies such as integrating random walks with the Word2Vec model \cite{perozzi2014deepwalk}, adjusting exploration and return variables of random walks \cite{grover2016node2vec}, leveraging second-order proximities between nodes \cite{tang2015line}, utilizing 3-motifs for distinguishing high-order graph structures \cite{chen2023motif}, or utilizing attention mechanisms \cite{velikovi2018graph} to train embedding vectors for all nodes in the network and ultimately validate classification performance based on the output layer using simple neural network structures like linear maps. To integrate our approach with these, we apply the softmax function to the $l$-dimensional output vector (where $l = |I|$ is the number of clusters or labels) of the existing architecture to create a probability distribution for each node. We then set these distributions as the initial parameters and use the proposed objective function to retrain. In other words, we use the output of the GNN architectures via softmax function instead of the \textsf{RW} to initialize our proposed optimization method \eqref{problem}. This strategy allows us to leverage ($l$-dimensional) classification vectors resulting from the trained embedding vectors of GNN-based architectures and use these classification vectors to learn the various higher-order structures included in the network using the proposed objective function, potentially improving classification performance. 
 
 \begin{table}[!t]
\centering
\includegraphics[width=3.5in]{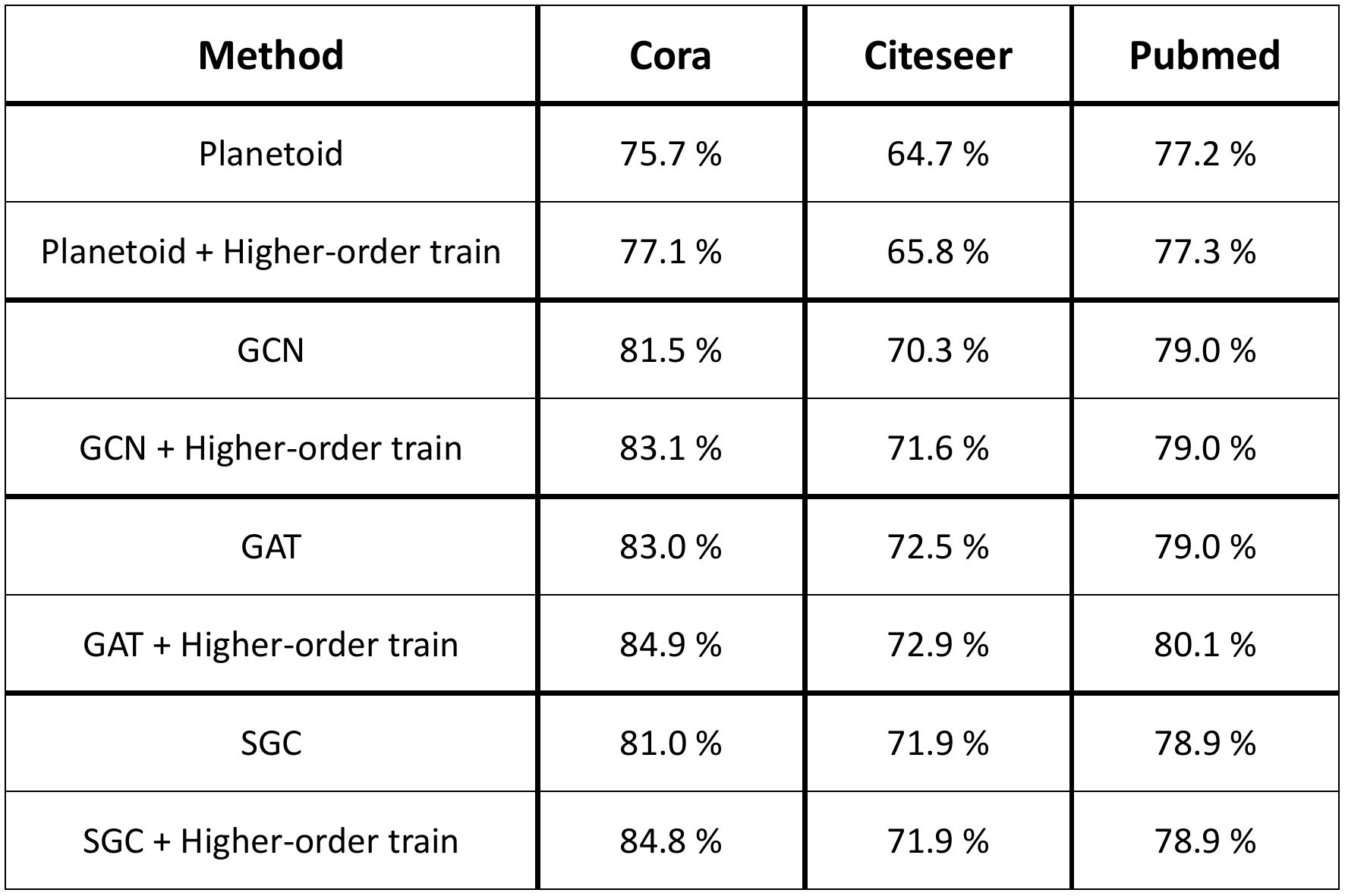}
\caption{Summary of accuracy results for benchmark datasets.}
\label{integrationtable}
\end{table}

In this experiment, we integrate GNNs with the proposed objective function and evaluate the performance gains using the Cora, Citeseer, and Pubmed datasets. GAT \cite{velikovi2018graph} uses an attention mechanism to learn node embeddings. The node features are created using the bag-of-words representation of documents, with the dimensions of the node features for Cora, Citeseer, and Pubmed being 1433, 3703, and 500, respectively. These features are used to learn embedding vectors using a multi-head attention structure, and a linear map is employed to generate outputs corresponding to the number of clusters. In our method, the total training and validation data used in \cite{velikovi2018graph}—640 for Cora, 620 for Citeseer, and 560 for Pubmed—are treated as the overall prior information. In contrast to \cite{velikovi2018graph}, which evaluated performance using 1,000 test nodes for each dataset, we assess performance on all remaining nodes not designated for training or validation. The treatment is similarly applied to Planetoid \cite{yang2016revisiting} (transductive experiment), GCN \cite{kipf2017semisupervised} (with 64 hidden units), and SGC \cite{pmlr-v97-wu19e}. Table \ref{integrationtable} summarizes the results of the integration experiments.

\subsection{Summary and discussion}\label{summary}

In this paper, we examined the performance of the proposed objective function \eqref{objective} in a variety of contexts, including balanced, imbalanced, and benchmark datasets. Several notable experimental results and discussions are listed below.

First, SBTM is a graph generation model that enables control over the distribution of $k$-cliques in the created network. Each element $(T_k)_\tta$ of a $k$-tensor $T_k \in \R^{l^k}$ reflects the probability of creating a $k$-clique given the label distribution $\tta$. We generated  $T_k$ systematically from a $l \times l$ matrices $B_k$ using the formula \eqref{probytensor} for each $k \ge 2$. The motivation is that by modifying the entries of $B_k$, one can effectively mimic networks with clique distributions that reflect real networks.

Second, it is clear that nodes corresponding to each label become more distinguishable in node classification tasks as the homo-connection probability increases, the hetero-connection probability decreases, and the prior information ratio rises, facilitating community detection. Given this, it is interesting that the proposed objective function showed significant performance gains in the opposite circumstance (lower diagonal, larger off-diagonal in $B_k$, and lower prior information ratio). This tendency is consistent across both balanced and imbalanced studies, showing that the suggested objective function can be applied in challenging classification scenarios, such as recognizing overlapping communities. 

Third, it is discovered that in the balanced \textsf{PI} experiment, the nodes representing incorrect predictions are uniformly distributed across all labels. In contrast, in the imbalanced \textsf{PI} experiment, nodes corresponding to incorrect predictions are weighted toward label indices with more nodes than other labels. Furthermore, experimental results show that the weight parameter $w_k=\alpha^{k-1}$ applied to $k$-cliques in the objective function \eqref{objective} yields that as $\alpha$ increases (meaning greater penalties are applied to higher-order simplices), additional performance gains are more pronounced in imbalanced networks than balanced ones. This suggests that the performance gain obtained with the proposed objective function, which is attributed to correcting the error distribution by using higher-order simplices in the training process, combined with the strategy of imposing larger penalties on higher-order simplices, could be more effective in imbalanced scenarios.

Fourth, the learning strategy used in this study does not rely on the many hyperparameters typically used in GNN-based algorithms, such as node embedding dimensions, the depth of the neural network architecture, the number of units in each layer, dropout ratios, or activation mechanisms. Additionally, experimental results on both synthetic networks and real citation networks have shown that the standard deviation of the results from each experiment using the proposed algorithm is negligible. This demonstrates that the proposed algorithm offers advantages in terms of consistency of results, stability, simplicity, and generalization performance. 

Fifth, as described in the objective \eqref{objective}, we use a strategy  where we first aggregate all $k$-cliques in a given network for each $k$, calculate the errors, and then repeat for all $k=2,3,…,M$. After that, we combine all of the errors and minimize the outcome. The multinomial coefficients ($C_\tta$ in \eqref{objective}) corresponding to the $k$-cliques differ for each $k$, making it difficult to apply a uniform coefficient across different $k$ values. This variability prevents standardizing the method for extracting the characteristics of cliques of different sizes in the network into a single method.  As a result, this characteristic of the objective function acts as a barrier to parallelization. We suggest that future research should focus on finding an effective method for gathering higher-order simplices in the network to enable parallel computing. 

Sixth, according to experimental results with benchmark data, using the training and validation data employed in Planetoid, GCN, GAT, and SGC as prior information improves mean accuracy by 2.2\%, 0.7\%, and 0.3\% for Cora, Citeseer, and Pubmed, respectively. The objective function used in this study is intended to promote that all nodes forming higher-order simplices exhibit similar node probability distributions. As a result, if the network has a large number of inter-simplices, where nodes producing higher-order simplices belong to distinct labels or clusters, achieving good performance with the proposed objective function becomes challenging. The proportions of inter-cluster simplices among all higher-order simplices beyond 2-simplices are 16.4\%, 18.8\%, and 22.6\% (therefore, intra-cluster simplices ratio would be 83.6\%, 81.2\%, 77.4\%) for Cora, Citeseer, and Pubmed, respectively. This observation explains why a higher accuracy improvement is observed with Cora compared to Citeseer and Pubmed. 

Seventh, in widely used real datasets such as Cora, Citeseer, and Pubmed, the number of labels $l$ is modest, allowing the proposed algorithm’s time complexity to remain manageable, and our method can be effectively applied without significant computational concerns. However, for datasets with a large number of labels, such as the Amazon Product Co-purchasing Network \cite{leskovec2016snap}, which contains hundreds or even thousands of labels, the computational cost may be too high. Because our algorithm's complexity increases exponentially with $l$, it may be impractical to utilize it on datasets with a large number of labels. This constraint highlights the need for algorithmic improvements to support applications with multiple labels.

Finally, this study demonstrated that the combination of GAT with the proposed objective function improved classification performance on benchmark data. While numerous recent advanced methodologies are being developed in the field of semi-supervised learning, almost all of them do not directly address higher-order simplices. If the given data supports the hypothesis of this study—densely connected nodes tend to exhibit similar attributes—it is speculated that further performance gains could be achieved by utilizing combinations with state-of-the-art approaches such as MGNN and GAT.

\section{Conclusion}\label{conclusion}

In this paper, we propose a probability-based objective function for semi-supervised node classification that takes advantage of simplicial interactions of varying order. Given that densely connected nodes are likely to have similar properties, our proposed objective function imposes a greater penalty when nodes connected via higher-order simplices have diversified labels. For a given number of distinct labels $l$, each node is equipped with an $l$-dimensional probability distribution, and we seek the distribution across all nodes that minimizes the objective function under the constraint that the sum of node probabilities is one. Furthermore, based on the recognition that traditional stochastic block models do not adequately mimic many real datasets, particularly in representing the distribution of higher-order simplices within each cluster, we propose the stochastic block tensor model (SBTM). For each $k \ge 2$, the SBTM uses probability parameters to control the number of $k$-cliques within or between clusters, adjusting the distribution of higher-order simplices in the network, thus better reflecting real data characteristics. The evaluation of our proposed function was conducted using graphs generated by the stochastic (SBTM) and in integration with graph neural network-based architectures (GAT). In challenging classification scenarios, where the probability of connections within the same label is low, the probability of connections between different labels is high, and there are fewer nodes with known labels, our proposed function integrating higher-order networks outperformed results from simple pairwise interactions or random walk-based probabilistic methods. Especially in imbalanced data, by adjusting the weight parameter within the objective function, further accuracy gains were achieved when the distribution of misclassified nodes was biased towards certain label indices containing more nodes than other labels. This offers potential applications in many node classification problems in network data, when combined with several semi-supervised studies that do not directly use higher-order simplices dispersed in networks. Our suggested objective function, which conducts different calculations depending on the size of the simplices, confronts a computational challenge because uniform operations cannot be applied to all simplices, making GPU-based parallel computing approaches difficult to deploy.  Overcoming this limitation is proposed for future research.

\section*{Acknowledgments}
Eunho Koo gratefully acknowledges the support of the Korea Institute
for Advanced Study (KIAS) under the individual grant AP086801. Tongseok Lim  wishes to express gratitude to the KIAS AI research group and director Hyeon, Changbong for their hospitality and support during his stay at KIAS in 2023 and 2024, where parts of this work were performed.


\bibliographystyle{IEEEtran}
\bibliography{NodeClassification}

\begin{thebibliography}{10}
\providecommand{\url}[1]{#1}
\csname url@samestyle\endcsname
\providecommand{\newblock}{\relax}
\providecommand{\bibinfo}[2]{#2}
\providecommand{\BIBentrySTDinterwordspacing}{\spaceskip=0pt\relax}
\providecommand{\BIBentryALTinterwordstretchfactor}{4}
\providecommand{\BIBentryALTinterwordspacing}{\spaceskip=\fontdimen2\font plus
\BIBentryALTinterwordstretchfactor\fontdimen3\font minus
  \fontdimen4\font\relax}
\providecommand{\BIBforeignlanguage}[2]{{%
\expandafter\ifx\csname l@#1\endcsname\relax
\typeout{** WARNING: IEEEtran.bst: No hyphenation pattern has been}%
\typeout{** loaded for the language `#1'. Using the pattern for}%
\typeout{** the default language instead.}%
\else
\language=\csname l@#1\endcsname
\fi
#2}}
\providecommand{\BIBdecl}{\relax}
\BIBdecl

\bibitem{fortunato2016community}
S.~Fortunato and D.~Hric, ``Community detection in networks: A user guide,''
  \emph{Physics reports}, vol. 659, pp. 1--44, 2016.

\bibitem{schaub2017many}
M.~T. Schaub, J.-C. Delvenne, M.~Rosvall, and R.~Lambiotte, ``The many facets
  of community detection in complex networks,'' \emph{Applied network science},
  vol.~2, no.~1, pp. 1--13, 2017.

\bibitem{easley2010networks}
D.~Easley, J.~Kleinberg \emph{et~al.}, \emph{Networks, crowds, and markets:
  Reasoning about a highly connected world}.\hskip 1em plus 0.5em minus
  0.4em\relax Cambridge university press Cambridge, 2010, vol.~1.

\bibitem{klamt2009hypergraphs}
S.~Klamt, U.-U. Haus, and F.~Theis, ``Hypergraphs and cellular networks,''
  \emph{PLoS computational biology}, vol.~5, no.~5, p. e1000385, 2009.

\bibitem{battiston2020networks}
F.~Battiston, G.~Cencetti, I.~Iacopini, V.~Latora, M.~Lucas, A.~Patania, J.-G.
  Young, and G.~Petri, ``Networks beyond pairwise interactions: Structure and
  dynamics,'' \emph{Physics Reports}, vol. 874, pp. 1--92, 2020.

\bibitem{holland1983stochastic}
P.~W. Holland, K.~B. Laskey, and S.~Leinhardt, ``Stochastic blockmodels: First
  steps,'' \emph{Social networks}, vol.~5, no.~2, pp. 109--137, 1983.

\bibitem{ghoshdastidar2014consistency}
D.~Ghoshdastidar and A.~Dukkipati, ``Consistency of spectral partitioning of
  uniform hypergraphs under planted partition model,'' \emph{Advances in Neural
  Information Processing Systems}, vol.~27, 2014.

\bibitem{kim2017community}
C.~Kim, A.~S. Bandeira, and M.~X. Goemans, ``Community detection in
  hypergraphs, spiked tensor models, and sum-of-squares,'' in \emph{2017
  International Conference on Sampling Theory and Applications (SampTA)}.\hskip
  1em plus 0.5em minus 0.4em\relax IEEE, 2017, pp. 124--128.

\bibitem{karrer2011stochastic}
B.~Karrer and M.~E. Newman, ``Stochastic blockmodels and community structure in
  networks,'' \emph{Physical review E}, vol.~83, no.~1, p. 016107, 2011.

\bibitem{airoldi2008mixed}
E.~M. Airoldi, D.~Blei, S.~Fienberg, and E.~Xing, ``Mixed membership stochastic
  blockmodels,'' \emph{Advances in neural information processing systems},
  vol.~21, 2008.

\bibitem{blondel2008fast}
V.~D. Blondel, J.-L. Guillaume, R.~Lambiotte, and E.~Lefebvre, ``Fast unfolding
  of communities in large networks,'' \emph{Journal of statistical mechanics:
  theory and experiment}, vol. 2008, no.~10, p. P10008, 2008.

\bibitem{newman2006modularity}
M.~E. Newman, ``Modularity and community structure in networks,''
  \emph{Proceedings of the national academy of sciences}, vol. 103, no.~23, pp.
  8577--8582, 2006.

\bibitem{vazquez2009finding}
A.~Vazquez, ``Finding hypergraph communities: a bayesian approach and
  variational solution,'' \emph{Journal of Statistical Mechanics: Theory and
  Experiment}, vol. 2009, no.~07, p. P07006, 2009.

\bibitem{chien2018community}
I.~Chien, C.-Y. Lin, and I.-H. Wang, ``Community detection in hypergraphs:
  Optimal statistical limit and efficient algorithms,'' in \emph{International
  Conference on Artificial Intelligence and Statistics}.\hskip 1em plus 0.5em
  minus 0.4em\relax PMLR, 2018, pp. 871--879.

\bibitem{ghoshdastidar2017consistency}
D.~Ghoshdastidar and A.~Dukkipati, ``Consistency of spectral hypergraph
  partitioning under planted partition model,'' 2017.

\bibitem{eaton2012spin}
E.~Eaton and R.~Mansbach, ``A spin-glass model for semi-supervised community
  detection,'' in \emph{Proceedings of the AAAI Conference on Artificial
  Intelligence}, vol.~26, no.~1, 2012, pp. 900--906.

\bibitem{ma2010semi}
X.~Ma, L.~Gao, X.~Yong, and L.~Fu, ``Semi-supervised clustering algorithm for
  community structure detection in complex networks,'' \emph{Physica A:
  Statistical Mechanics and its Applications}, vol. 389, no.~1, pp. 187--197,
  2010.

\bibitem{liu2014semi}
D.~Liu, X.~Liu, W.~Wang, and H.~Bai, ``Semi-supervised community detection
  based on discrete potential theory,'' \emph{Physica A: Statistical Mechanics
  and its Applications}, vol. 416, pp. 173--182, 2014.

\bibitem{nowicki2001estimation}
K.~Nowicki and T.~A.~B. Snijders, ``Estimation and prediction for stochastic
  blockstructures,'' \emph{Journal of the American statistical association},
  vol.~96, no. 455, pp. 1077--1087, 2001.

\bibitem{kipf2017semisupervised}
T.~N. Kipf and M.~Welling, ``Semi-supervised classification with graph
  convolutional networks,'' in \emph{International Conference on Learning
  Representations}, 2017.

\bibitem{hamilton2017inductive}
W.~Hamilton, Z.~Ying, and J.~Leskovec, ``Inductive representation learning on
  large graphs,'' \emph{Advances in neural information processing systems},
  vol.~30, 2017.

\bibitem{velikovi2018graph}
P.~Veli{\v c}kovi{\'c}, G.~Cucurull, A.~Casanova, A.~Romero, P.~Li{\`o}, and
  Y.~Bengio, ``Graph attention networks,'' \emph{International Conference on
  Learning Representations}, 2018.

\bibitem{xu2018representation}
K.~Xu, C.~Li, Y.~Tian, T.~Sonobe, K.-i. Kawarabayashi, and S.~Jegelka,
  ``Representation learning on graphs with jumping knowledge networks,'' in
  \emph{International conference on machine learning}.\hskip 1em plus 0.5em
  minus 0.4em\relax PMLR, 2018, pp. 5453--5462.

\bibitem{chen2023motif}
X.~Chen, R.~Cai, Y.~Fang, M.~Wu, Z.~Li, and Z.~Hao, ``Motif graph neural
  network,'' \emph{IEEE Transactions on Neural Networks and Learning Systems},
  2023.

\bibitem{bendito2003solving}
E.~Bendito, A.~Carmona, and A.~M. Encinas, ``Solving dirichlet and poisson
  problems on graphs by means of equilibrium measures,'' \emph{European Journal
  of Combinatorics}, vol.~24, no.~4, pp. 365--375, 2003.

\bibitem{bick2023higher}
C.~Bick, E.~Gross, H.~A. Harrington, and M.~T. Schaub, ``What are higher-order
  networks?'' \emph{SIAM Review}, vol.~65, no.~3, pp. 686--731, 2023.

\bibitem{lim2020hodge}
L.-H. Lim, ``Hodge laplacians on graphs,'' \emph{Siam Review}, vol.~62, no.~3,
  pp. 685--715, 2020.

\bibitem{paul2023higher}
S.~Paul, O.~Milenkovic, and Y.~Chen, ``Higher-order spectral clustering under
  superimposed stochastic block models,'' \emph{Journal of Machine Learning
  Research}, vol.~24, no. 320, pp. 1--58, 2023.

\bibitem{song2022graph}
Z.~Song, X.~Yang, Z.~Xu, and I.~King, ``Graph-based semi-supervised learning: A
  comprehensive review,'' \emph{IEEE Transactions on Neural Networks and
  Learning Systems}, vol.~34, no.~11, pp. 8174--8194, 2022.

\bibitem{zhu2003semi}
X.~Zhu, Z.~Ghahramani, and J.~D. Lafferty, ``Semi-supervised learning using
  gaussian fields and harmonic functions,'' in \emph{Proceedings of the 20th
  International conference on Machine learning (ICML-03)}, 2003, pp. 912--919.

\bibitem{zhou2003learning}
D.~Zhou, O.~Bousquet, T.~Lal, J.~Weston, and B.~Sch{\"o}lkopf, ``Learning with
  local and global consistency,'' \emph{Advances in neural information
  processing systems}, vol.~16, 2003.

\bibitem{zhou2005learning}
D.~Zhou, J.~Huang, and B.~Sch{\"o}lkopf, ``Learning from labeled and unlabeled
  data on a directed graph,'' in \emph{Proceedings of the 22nd international
  conference on Machine learning}, 2005, pp. 1036--1043.

\bibitem{wang2006label}
F.~Wang and C.~Zhang, ``Label propagation through linear neighborhoods,'' in
  \emph{Proceedings of the 23rd international conference on Machine learning},
  2006, pp. 985--992.

\bibitem{belkin2006manifold}
M.~Belkin, P.~Niyogi, and V.~Sindhwani, ``Manifold regularization: A geometric
  framework for learning from labeled and unlabeled examples.'' \emph{Journal
  of machine learning research}, vol.~7, no.~11, 2006.

\bibitem{roweis2000nonlinear}
S.~T. Roweis and L.~K. Saul, ``Nonlinear dimensionality reduction by locally
  linear embedding,'' \emph{science}, vol. 290, no. 5500, pp. 2323--2326, 2000.

\bibitem{belkin2001laplacian}
M.~Belkin and P.~Niyogi, ``Laplacian eigenmaps and spectral techniques for
  embedding and clustering,'' \emph{Advances in neural information processing
  systems}, vol.~14, 2001.

\bibitem{cao2015grarep}
S.~Cao, W.~Lu, and Q.~Xu, ``Grarep: Learning graph representations with global
  structural information,'' in \emph{Proceedings of the 24th ACM international
  on conference on information and knowledge management}, 2015, pp. 891--900.

\bibitem{ou2016asymmetric}
M.~Ou, P.~Cui, J.~Pei, Z.~Zhang, and W.~Zhu, ``Asymmetric transitivity
  preserving graph embedding,'' in \emph{Proceedings of the 22nd ACM SIGKDD
  international conference on Knowledge discovery and data mining}, 2016, pp.
  1105--1114.

\bibitem{perozzi2014deepwalk}
B.~Perozzi, R.~Al-Rfou, and S.~Skiena, ``Deepwalk: Online learning of social
  representations,'' in \emph{Proceedings of the 20th ACM SIGKDD international
  conference on Knowledge discovery and data mining}, 2014, pp. 701--710.

\bibitem{tang2015line}
J.~Tang, M.~Qu, M.~Wang, M.~Zhang, J.~Yan, and Q.~Mei, ``Line: Large-scale
  information network embedding,'' in \emph{Proceedings of the 24th
  international conference on world wide web}, 2015, pp. 1067--1077.

\bibitem{yang2016revisiting}
Z.~Yang, W.~Cohen, and R.~Salakhudinov, ``Revisiting semi-supervised learning
  with graph embeddings,'' in \emph{International conference on machine
  learning}.\hskip 1em plus 0.5em minus 0.4em\relax PMLR, 2016, pp. 40--48.

\bibitem{grover2016node2vec}
A.~Grover and J.~Leskovec, ``node2vec: Scalable feature learning for
  networks,'' in \emph{Proceedings of the 22nd ACM SIGKDD international
  conference on Knowledge discovery and data mining}, 2016.

\bibitem{hong2021variational}
X.~Hong, T.~Zhang, Z.~Cui, and J.~Yang, ``Variational gridded graph convolution
  network for node classification,'' \emph{IEEE/CAA Journal of Automatica
  Sinica}, vol.~8, no.~10, pp. 1697--1708, 2021.

\bibitem{wang2016structural}
D.~Wang, P.~Cui, and W.~Zhu, ``Structural deep network embedding,'' in
  \emph{Proceedings of the 22nd ACM SIGKDD international conference on
  Knowledge discovery and data mining}, 2016, pp. 1225--1234.

\bibitem{cao2016deep}
S.~Cao, W.~Lu, and Q.~Xu, ``Deep neural networks for learning graph
  representations,'' in \emph{Proceedings of the AAAI conference on artificial
  intelligence}, vol.~30, no.~1, 2016.

\bibitem{kipf2016variational}
T.~N. Kipf and M.~Welling, ``Variational graph auto-encoders,'' \emph{NIPS
  Workshop on Bayesian Deep Learning}, 2016.

\bibitem{wang2022minority}
K.~Wang, J.~An, M.~Zhou, Z.~Shi, X.~Shi, and Q.~Kang, ``Minority-weighted graph
  neural network for imbalanced node classification in social networks of
  internet of people,'' \emph{IEEE Internet of Things Journal}, vol.~10, no.~1,
  pp. 330--340, 2022.

\bibitem{scarselli2008graph}
F.~Scarselli, M.~Gori, A.~C. Tsoi, M.~Hagenbuchner, and G.~Monfardini, ``The
  graph neural network model,'' \emph{IEEE transactions on neural networks},
  vol.~20, no.~1, pp. 61--80, 2008.

\bibitem{zaheer2017deep}
M.~Zaheer, S.~Kottur, S.~Ravanbakhsh, B.~Poczos, R.~R. Salakhutdinov, and A.~J.
  Smola, ``Deep sets,'' \emph{Advances in neural information processing
  systems}, vol.~30, 2017.

\bibitem{pmlr-v97-wu19e}
F.~Wu, A.~Souza, T.~Zhang, C.~Fifty, T.~Yu, and K.~Weinberger, ``Simplifying
  graph convolutional networks,'' in \emph{Proceedings of the 36th
  International Conference on Machine Learning}, vol.~97.\hskip 1em plus 0.5em
  minus 0.4em\relax PMLR, 2019.

\bibitem{WangYWC0ZHC19}
L.~Wang, W.~Yu, W.~Wang, W.~Cheng, W.~Zhang, H.~Zha, X.~He, and H.~Chen,
  ``Learning robust representations with graph denoising policy network,'' in
  \emph{{IEEE} International Conference on Data Mining}, 2019.

\bibitem{SunHV020}
F.~Sun, J.~Hoffmann, V.~Verma, and J.~Tang, ``Infograph: Unsupervised and
  semi-supervised graph-level representation learning via mutual information
  maximization,'' in \emph{8th International Conference on Learning
  Representations, {ICLR} 2020}.

\bibitem{pmlr-v119-zheng20d}
C.~Zheng, B.~Zong, W.~Cheng, D.~Song, J.~Ni, W.~Yu, H.~Chen, and W.~Wang,
  ``Robust graph representation learning via neural sparsification,'' in
  \emph{Proceedings of the 37th International Conference on Machine Learning},
  2020.

\bibitem{10.1145/3437963.3441734}
D.~Luo, W.~Cheng, W.~Yu, B.~Zong, J.~Ni, H.~Chen, and X.~Zhang, ``Learning to
  drop: Robust graph neural network via topological denoising,'' in
  \emph{Proceedings of the 14th ACM International Conference on Web Search and
  Data Mining}, 2021.

\bibitem{gong2022self}
M.~Gong, H.~Zhou, A.~K. Qin, W.~Liu, and Z.~Zhao, ``Self-paced co-training of
  graph neural networks for semi-supervised node classification,'' \emph{IEEE
  Transactions on Neural Networks and Learning Systems}, vol.~34, no.~11, pp.
  9234--9247, 2022.

\bibitem{sen2008collective}
P.~Sen, G.~Namata, M.~Bilgic, L.~Getoor, B.~Galligher, and T.~Eliassi-Rad,
  ``Collective classification in network data,'' \emph{AI magazine}, vol.~29,
  no.~3, pp. 93--93, 2008.

\bibitem{glorot2010understanding}
X.~Glorot and Y.~Bengio, ``Understanding the difficulty of training deep
  feedforward neural networks,'' in \emph{Proceedings of the thirteenth
  international conference on artificial intelligence and statistics}.\hskip
  1em plus 0.5em minus 0.4em\relax JMLR Workshop and Conference Proceedings,
  2010, pp. 249--256.

\bibitem{KingmaB14}
D.~P. Kingma and J.~Ba, ``Adam: {A} method for stochastic optimization,'' in
  \emph{3rd International Conference on Learning Representations}, 2015.

\bibitem{leskovec2016snap}
J.~Leskovec and R.~Sosi{\v{c}}, ``Snap: A general-purpose network analysis and
  graph-mining library,'' \emph{ACM Transactions on Intelligent Systems and
  Technology (TIST)}, vol.~8, no.~1, pp. 1--20, 2016.

\end{thebibliography}


\begin{IEEEbiography}[{\includegraphics[width=1in,height=1.25in,clip,keepaspectratio]{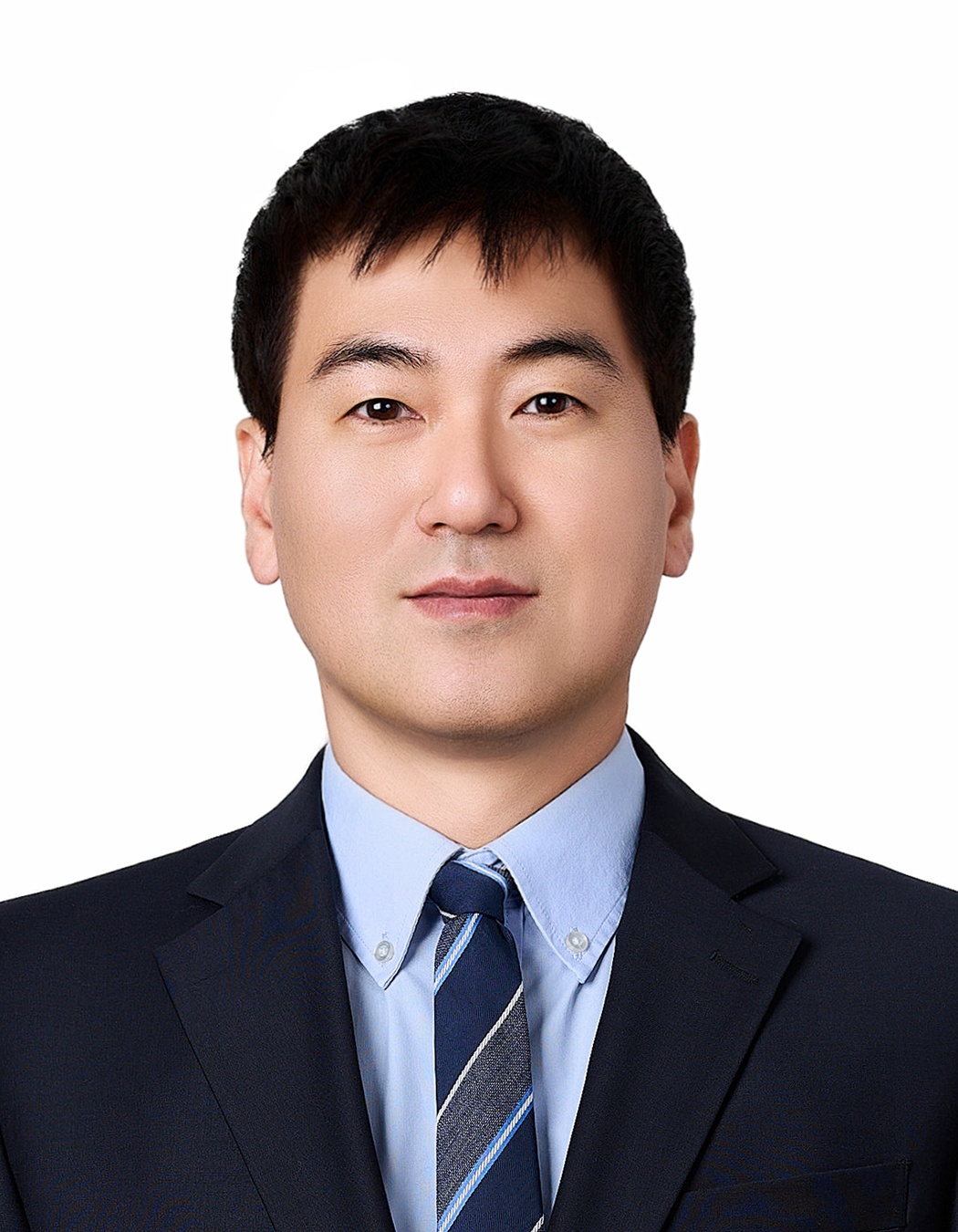}}]{Eunho Koo}
received his Ph.D. in Applied Mathematics from Yonsei University, South Korea, in 2017. He worked as a researcher at Institute of Industrial Science, University of Tokyo, Japan, and  as a researcher at Center for AI and Natural Sciences, Korea Institute for Advanced Study, South Korea.
He is currently working as an associate professor at Department of Big Data Convergence, Chonnam National University, South Korea. 
 His research interests include statistical machine learning and prediction of time series based on neural networks.
\end{IEEEbiography}

\begin{IEEEbiography}[{\includegraphics[width=1in,height=1.25in,clip,keepaspectratio]{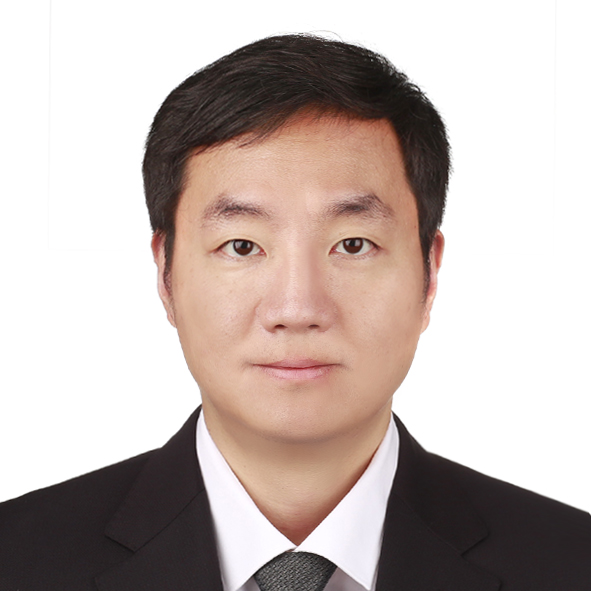}}]{Tongseok Lim} earned his Ph.D. in Mathematics from The University of British Columbia, Canada, in 2016. He is currently an assistant professor in the Quantitative Methods Department at Purdue University's Daniels School of Business. Previously, he served as an assistant professor at ShanghaiTech University and held postdoctoral research positions at the University of Oxford and the Vienna University of Technology. His research focuses on Variational Methods and Optimal Transport, with its applications in Economics, Finance, and Data Science.
\end{IEEEbiography}

\end{document}